\documentclass[11pt]{article}

\usepackage[latin1]{inputenc}
\usepackage{tikz}
\usetikzlibrary{trees}
\usepackage{verbatim}

\usepackage{amsfonts}
\usepackage{amssymb}
\usepackage{amsmath}
\usepackage{amsthm}
\usepackage{dsfont}
\usepackage{bbm}
\usepackage{graphics}
\usepackage{graphicx}
\usepackage{epsfig}
\usepackage{mathrsfs}
\usepackage[all]{xy}
\usepackage{color}

\newtheorem{theorem}{Theorem}[section]

\newtheorem{definition}[theorem]{Definition}

\newtheorem{lemma}[theorem]{Lemma}
\newtheorem{proposition}[theorem]{Proposition}
\newtheorem{remark}[theorem]{Remark}

\numberwithin{equation}{section}

\def\id{\mathbf{1}}
\def\11{\mathbbold{1}}
\def\qed{{\hbox{ }\hfill {$\Box$}}}
\def\apru{{\,\underset{u}\approx}\,}
\def\uapr{{\,\overset{u}\approx}\,}
\def\trieq{\triangleq}

\def\proof{\noindent\textbf{Proof. }}

\def\intr{\mathrm{int}}

\def\cone{\mathrm{cone}}

\def\Rbb{\mathbb{R}}

\def\Fc{\mathcal{F}}

\def\Ic{\mathcal{I}}

\def\Bsc{\mathscr{B}}
\def\Csc{\mathscr{C}}

\def\Jsc{\mathscr{J}}
\def\Ksc{\mathscr{K}}

\def\Psc{\mathscr{P}}
\def\Qsc{\mathscr{Q}}

\title{Decision Making under Uncertainty:\\ A Game of Two Selves\thanks{Supported by National Key R\&D Program of China (NO. 2018YFA0703900).}}
\author{Jianming Xia\thanks{RCSDS, NCMIS, Academy of Mathematics and Systems Science,
Chinese Academy of Sciences, Beijing 100190, China; Email:
xia@amss.ac.cn.}
}
\date{}

\begin{document}

\maketitle

\begin{abstract}
In this paper we characterize the niveloidal preferences that satisfy the Weak Order, Monotonicity, Archimedean, and Weak C-Independence Axioms from the point of view of an intra-personal, leader-follower game. We also show that the leader's strategy space can serve as an ambiguity aversion index.

\end{abstract}

\section{Introduction}

In order to resolve the paradox observed by Ellsberg (1961) and
violated by the subjective expected utility (SEU) of
Savage (1954) and Anscombe and Aumann (1963), Gilboa and Schmeidler
(1989) introduced the maxmin expected utility (maxmin EU), with which the decision maker ranks Anscombe-Aumann acts according to the minimal expected utility over a set of probabilities.
Motivated by the robust control approach, Hansen and Sargent (2000,
2001) proposed the multiplier preference, which was axiomatized by Strzalecki (2011).
Maccheroni, Marinacci and Rustichini (2006), MMR (2006) henceforth, introduced the variational
preference (VP), which includes as special cases the maxmin EU and the multiplier preference.
Chateauneuf and Faro (2009) introduced the confidence preferences.
More recently, Cerreia-Vioglio, Maccheroni, Marinacci and Montrucchio (2011), C3M (2011) henceforth, studied the
uncertainty averse preference (UAP), which includes  as special cases the VP and the confidence preference. 
All of the above preferences satisfy the ``uncertainty aversion" axiom. 

On the other hand, Schmeidler (1989) introduced the Choquet expected utility (CEU), with which the decision maker ranks
acts according to the Choquet expected utility with respect to a capacity (also called nonadditive probability). 
Generally, CEU preferences are neither uncertainty averse nor uncertainty seeking. 
As pointed out by Schmeidler (1989), a CEU preference is uncertainty averse if and only if the corresponding capacity is convex.  
Moreover, later studies show that people's attitude toward ambiguity is not systematically negative. For instance, a series of experiments conducted by Heath and Tversky (1991) show that agents can even be ambiguity-seeking when they consider themselves knowledgeable or competent. 
Other non uncertainty averse preferences include the $\alpha$-maxmin expected utility ($\alpha$-MEU)\footnote{The $\alpha$-MEU was axiomatized by GMM (2004). }
 (Hurwicz 1951, Arrow and Hurwicz 1972), the invariant biseparable preference (IB preference) of Ghirardato, Maccheroni and Marinacci (2004), GMM (2004) henceforth, and the smooth ambiguity preference of Klibanoff, Marinacci and Mukerji (2005), KMM (2005) henceforth. 

The IB preference was firstly characterized as the generalized $\alpha$-MEU preference by GMM (2004), where $\alpha$ depends on the act. Such kind of generalized $\alpha$-MEU characterization was further extended to MBA preferences (for Monotonic, Bernoullian, and Archimedean) by Cerreia-Vioglio,  Ghirardato, Maccheroni, Marinacci and Siniscalchi (2011), CGMMS (2011) henceforth. The IB preference was also characterized as the second-order CEU by Giraud (2005) and Amarante (2009). 

It is a natural idea to imagine that a decision maker with a non uncertainty averse preference is composed of two selves: 
uncertainty averse self and uncertainty seeking self, and the interaction between the two selves yields the decision maker's final evaluation of the act. This can be inspired by a closer looking at the $\alpha$-MEU. It is clear that, for any act $f$, 
\begin{subequations}\label{eq:alpha:meu0}
\begin{align}
&\alpha\min_{p\in P}\int u(f)\,dp+(1-\alpha)\max_{p\in P}\int u(f)\,dp\nonumber\\
=&\max_{p_1\in P} \min_{p_2\in P}\int u(f)d(\alpha p_1+(1-\alpha)p_2)\label{eq:alpha:meu:0a}\\
=&\min_{p_2\in P} \max_{p_1\in P}\int u(f)d(\alpha p_1+(1-\alpha)p_2)\label{eq:alpha:meu:0b},
\end{align}
\end{subequations}
where $P$ is convex and compact set consisting some probabilities and $\alpha\in[0,1]$. The reformulation \eqref{eq:alpha:meu0} provides an interpretation of $\alpha$-MEU from the point of view of a zero-sum game.  Consider a zero-sum game as follows. Players 1 and 2 choose priors $p_1$ and $p_2$ from $P$, respectively. A common probability is then given by the $\alpha$-combination $\alpha p_1+(1-\alpha)p_2$, under which the expected utility is calculated. 
In this game, player 1 acts as the uncertainty seeking self and player 2 the uncertainty averse self.  Then players 1 and 2 evaluate the act according to the maxmax EU and the maxmin EU, respectively. As a result, the act $f$ is finally evaluated by the equilibrium value (saddle point value) of the game. Therefore, the $\alpha$-MEU can be interpreted as the consequence of the interaction between the two selves and the zero-sum game between the two selves can be regarded as an intra-personal game that represents the mechanism of the interaction. 

The above interpretation from the point of view of a zero-sum game can be extended to the generalized $\alpha$-MEU characterization of the IB preference (respectively, MBA preferences). The only difference is that, for these preferences, the combination weight $\alpha$ is not a given constant but a function of the act $f$ (respectively, the state-wise utility $u\circ f$). Therefore, the generalized $\alpha$-MEU can be regarded as the equilibrium value (saddle point value) of a zero-sum game. 

This paper provides another interpretation from the point view of a leader-follower game. 
It turns out in Section \ref{sec:ibp} that the decision maker with the IB preference ranks the act $f$ according to 
\begin{subequations}\label{eq:ibp:0}
\begin{align}
V(f)=&\max_{P\in\Psc} \min_{p\in P}\int u(f)\,dp\label{eq:ibp:0a}\\
=&\min_{Q\in\Qsc} \max_{q\in Q}\int u(f)\,dq,\label{eq:ibp:0b}
\end{align}
\end{subequations}
where $\Psc$ and $\Qsc$ are two families consisting of some convex and compact subsets $P,Q\subseteq\Delta$ and $\Delta$ is the set consisting of all probabilities. The characterization \eqref{eq:ibp:0} provides an interpretation of the IB preference from the point of view of a leader-follower game. For instance, in representation \eqref{eq:ibp:0a}, the leader acts as the uncertainty seeking self and the follower the uncertainty averse self.
The Stackelerg equilibrium value of the game yields the final evaluation of the act. Moreover, such characterization by an intra-personal, leader-follower game can be extended to the more general preference (called the niveloidal preference in this paper),  which satisfies the Weak C-Independence Axiom as well as the Weak Order, Monotonicity, and Archimedean Axioms. 

The main contribution of this paper is to characterize the niveloidal preferences from the point of view of an intra-personal, leader-follower game. Another contribution of this paper is to show that the leader's strategy space can serve as an ambiguity aversion index. For instance, in representation \eqref{eq:ibp:0a}, $\Psc$ is the strategy space of the leader, who acts as the uncertainty seeking self; the lager $\Psc$ is, the less ambiguity averse the decision maker is. In representation \eqref{eq:ibp:0b}, $\Qsc$ is the strategy space of the leader, who now acts as the uncertainty averse self; the lager $\Qsc$ is, the more ambiguity averse the decision maker is. The level of ambiguity aversion is finally determined by the leader!

The rest part of this paper is organized as follows. The set up and
some mathematical preliminaries are introduced in Section
\ref{sec:pre}. The representation result and interpretations for niveloidal preferences are presented in Section \ref{sec:nive}. The representation result and interpretations for IB preferences are presented in Section \ref{sec:ibp}. Some special preferences are investigated in Section \ref{sec:special}. The ambiguity attitude is discussed in Section \ref{sec:amg}. Proofs
and related material are collected in the Appendices.

\section{Preliminaries}\label{sec:pre}

Let $S$ be a nonempty set of \textit{states of world}, $\Sigma$ an
algebra of subsets of $S$ called \textit{events}, and $X$ a nonempty
set of \textit{consequences}. We denoted by $\Fc$ the set of all the
\textit{simple acts}: mappings $f:S\rightarrow X$ which have
finitely many values and are $\Sigma$-measurable. 
It is customary to identify every consequence $x\in X$ with the
constant act $x\in\Fc$ that yields $x$ in every state. 

We always assume $X$ is a convex subset of a vector space. For example,
this is the case if $X$ is the set of all 
distributions on a nonempty set $Z$ with finite support:
\begin{align*}
X=\left\{x:Z\rightarrow[0,1]\,\left|\, x(z)\ne0\mbox{ for finitely many }z\in Z\mbox{ and }
\sum_{z\in Z}x(z)=1\right.\right\}.
\end{align*}
For every $f,g\in\Fc$ and $\alpha\in[0,1]$, we can define the mixed
act $\alpha f+(1-\alpha)g$ which yields consequence $\alpha
f(s)+(1-\alpha)g(s)$ in every state $s$.

Let $B_0(\Sigma)$ be the set of all $\Rbb$-valued
$\Sigma$-measurable functions that have finitely many values and
$B(\Sigma)$ its supnorm closure. When endowed with the supnorm,
$B_0(\Sigma)$ is a normed vector space and $B(\Sigma)$ is a Banach
space. For every event $A\in\Sigma$, we use $\id_A$ to denote the
indicator function of $A$. As usual, for every $\alpha\in\Rbb$,
$\alpha \id_S$ is identified with $\alpha$. Obviously, $B_0(\Sigma)$
is the vector space spanned by $\{\id_A\,|\, A\in\Sigma\}$. For any
interval $T\subseteq\Rbb$, let $B_0(\Sigma,T)$ (resp. $B(\Sigma,T)$)
denote the set of all functions in $B_0(\Sigma)$ (resp. $B(\Sigma)$)
taking values in $T$. 
For every act $f\in\Fc$ and function $u:X\rightarrow\Rbb$, $u(f)$ is the element of
$B_0(\Sigma)$ defined by $u(f)(s)=u(f(s))$ for all $s\in S$.

The norm dual of $B_0(\Sigma)$ (resp. $B(\Sigma)$) is the space $ba(\Sigma)$ of all bounded
and finitely additive set functions on $\Sigma$ endowed with the total variation norm, the duality being
$\langle\varphi,m\rangle=\int\varphi\,dm$ for all $\varphi\in B_0(\Sigma)$ (resp. $B(\Sigma)$) and all $m\in ba(\Sigma)$;
see, e.g., Dundorf and Schwartz (1958, p. 258).
We write indifferently $\int\varphi\, dm$ or $m(\varphi)$. A nonnegative
element of $ba(\Sigma)$ that assigns value $1$ to $S$ is called a
\textit{finitely additive probability}, and it is usually denoted by $p$.
The set of all finitely additive probabilities on $\Sigma$ is denoted by $\Delta$.
Let $\Delta$ be endowed with the weak* topology, i.e., the $\sigma(\Delta, B_0(\Sigma))$ topology.

Let $\Phi$ be a nonempty subset of $B(\Sigma)$. A function
$I: \Phi\rightarrow[-\infty,\infty]$ is called:
\begin{itemize}
  \item \textit{normalized} if $I(\alpha)=\alpha$ for all $\alpha\in\Rbb\cap\Phi$;

  \item \textit{monotone} if $I(\varphi)\ge I(\psi)$ for all $\varphi,\psi\in \Phi$
  such that $\varphi\ge \psi$;

  \item \textit{translation invariant} if $I(\varphi+\alpha)=I(\varphi)+\alpha$ for all
  $\varphi\in \Phi$ and $\alpha\in\Rbb$ with $\varphi+\alpha\in\Phi$.
\end{itemize}

We recall the following definition of Dolecki and Greco (1995); see also MMR (2006).

\begin{definition}\label{def:novelloid}
Let $\Phi$ be a nonempty subset of $B(\Sigma)$. A function
$I: \Phi\rightarrow\Rbb$ is a \textit{niveloid} if it is monotone and translation invariant.
\end{definition}

\section{Niveloidal Preferences}\label{sec:nive}

The decision maker's preference is given by a binary relation $\succsim$ on
$\Fc$. As usual, its symmetric and asymmetric parts are denoted by
$\sim$ and $\succ$ respectively.

\subsection{Representations}

We first list the axioms that the preference satisfies.

\begin{description}
\item[Axiom A1 (Weak Order)] \textit{The binary relation $\succsim$
is nontrivial, complete, and transitive on $\Fc$.}

\item[Axiom A2 (Monotonicity)] \textit{For any $f,g\in \Fc$, if
$f(s)\succsim g(s)$ for all $s\in S$, then $f\succsim g$.}

\item[Axiom A3 (Archimedean)] \textit{For all $f,g,h\in\Fc$, if
$f\succ g\succ h$ then there exist some
$\alpha,\beta\in(0,1)$ such that $\alpha f+(1-\alpha)h\succ g\succ \beta f+(1-\beta)h$.}

\item[Axiom A4 (Weak C-Independence)] \textit{For all $f,g\in\Fc$,
$x,y\in X$, and $\alpha\in(0,1)$,
$$\alpha f+(1-\alpha)x\succsim  \alpha g+(1-\alpha)x\Rightarrow
\alpha f+(1-\alpha)y\succsim\alpha g+(1-\alpha)y.$$}
\end{description}

Axioms A1--A3 are standard and well understood. In Axiom A1,
nontriviality means that $f\succ g$ for some $f, g\in\Fc$.
Axiom A4 was introduced by MMR (2006) in their characterization of Variational Preferences.

The following representation result is from MMR (2006, Lemma 28).

\begin{lemma}\label{lma:niveloid}
A binary relation $\succsim$ on $\Fc$ satisfies axioms A1--A4 if and only if there exists a nonconstant affine function $u: X\to\Rbb$, with $0\in\mathrm{int}(u(X))$, and a normalized niveloid $I: B_0(\Sigma, u(X))\to\Rbb$ such that, for any $f,g\in\Fc$,
\begin{equation*}
f\succsim g\Leftrightarrow I(u(f))\ge I(u(g)).
\end{equation*}
Moreover,  $u$ is cardinally unique and, given $u$, $I$ is unique.
\end{lemma}

In accounting for the previous representation result, we give the following definition.

\begin{definition}\label{def:niveloidal}
A binary relation $\succsim$ is a niveloidal preference if it satisfies Axioms A1--A4.
\end{definition}

Let $\Csc$ denote the set of all lower semi-continuous and convex functions $c:\Delta\to(-\infty,\infty]$.
We need the following definition.

\begin{definition} A subset $C\subseteq\Csc$ is called \textit{grounded} if 
$\sup_{c\in C}\min_{p\in\Delta}c(p)=0$.
\end{definition}

In the case of $C=\{c\}$ being a singleton,  $C$ is grounded if and only if $\min_{p\in\Delta}c(p)=0$. Therefore, it is consistent with the definition of the groundedness for a single function $c$ in MMR (2006).

The next theorem characterizes the niveloidal preferences.

\begin{theorem}\label{thm:nive:rep}
Let $\succsim$ be a binary relation on $\Fc$. The following conditions are equivalent:

\begin{description}
  \item[(i)] $\succsim$ is a niveloidal preference;

  \item[(ii)] There exists a nonconstant affine function $u: X\rightarrow\Rbb$ and a grounded subset $C\subseteq\Csc$ such that,  for any $f$ and $g$ in $\Fc$,
\begin{align}\label{nive:eq-fg-maxmin}
f\succsim g\Leftrightarrow
\max_{c\in C}\min_{p\in\Delta}\left(\int u(f)\,dp+c(p)\right)
\ge\max_{c\in C}\min_{p\in\Delta}\left(\int u(g)\,dp+c(p)\right);
\end{align}

\item[(iii)] There exists a nonconstant affine function $u: X\rightarrow\Rbb$ and a grounded subset $B\subseteq\Csc$ such that, for any $f$ and $g$ in $\Fc$,
\begin{align}\label{nive:eq-fg-minmax}
f\succsim g\Leftrightarrow
\min_{b\in B}\max_{q\in\Delta}\left(\int u(f)\,dq-b(q)\right)
\ge\min_{b\in B}\max_{q\in\Delta}\left(\int u(g)\,dq-b(q)\right).
\end{align}
\end{description}

The function $u$ is cardinally unique, there is a
(unique) maximal grounded subset $C^*\subseteq\Csc$ satisfying
\eqref{nive:eq-fg-maxmin}, given by
\begin{align}\label{eq:c*}
C^*=\left\{c\in\Csc\,\left|\, \min_{p\in \Delta}\left(\int u(f)\,dp+c(p)\right)\le u(x_f)\mbox{ for all }f\in\Fc\right.\right\},
\end{align}
and there is a
(unique) maximal grounded subset $B^*\subseteq\Csc$ satisfying
\eqref{nive:eq-fg-minmax}, given by
\begin{align}\label{eq:b*}
B^*=\left\{b\in\Csc\,\left|\, \max_{p\in \Delta}\left(\int u(f)\,dq-b(q)\right)\ge u(x_f)\mbox{ for all }f\in\Fc\right.\right\}.
\end{align}
\end{theorem}

\proof See Appendix \ref{sec:proofs}. \qed

\subsection{Interpretations: Leader-Follower Games}

A decision maker with a niveloidal preference ranks acts $f$ according to the preference functional
\begin{align}\label{eq:vf:nive:p}
V(f)=\max_{c\in C}\min_{p\in \Delta}\left(\int u(f)\,dp+c(p)\right)
\end{align}
or
\begin{align}\label{eq:vf:nive:q}
V(f)=\min_{b\in B}\max_{q\in\Delta}\left(\int u(f)\,dq-b(q)\right).
\end{align}

Each of the above two representations can be related to a leader-follower game between two selves of the decision maker. More precisely, the decision maker is composed of two selves: uncertainty averse self and uncertainty seeking self. 
We first consider the game related to representation \eqref{eq:vf:nive:p}, where the uncertainty seeking self is the leader and the uncertainty averse self is the follower.
The game has two stages.
\begin{description}
\item[Stage 1:] The leader moves first and chooses a strategy $c\in C$.
\item[Stage 2:] Then in response, the follower (the uncertainty averse self) moves and chooses a strategy $p\in\Delta$ to minimize $\left(\int u(f)\,dp+c(p)\right)$, acting according to the uncertainty averse variational preference with penalty function $c$. Here an \textit{uncertainty averse variational preference}\footnote{Variational preferences (preferences that satisfy Axioms A1--A5) are investigated in MMR (2006). They are called uncertainty averse variational preferences here in order to be distinguished from those that satisfying Axioms A1--A4 and A6, which are called uncertainty seeking variational preferences here.  A similar way of MMR (2006) leads to the following choice criterion of an uncertainty seeking variational preference:
$$\max_{q\in\Delta}\left(\int u(f)\,dq-b(q)\right),$$
where $b\in\Csc$.} refers to a preference that satisfies Axioms A1--A4 and the next one:
\item[Axiom A5 (Uncertainty Aversion)] For all $f,g\in\Fc$ and $\alpha\in(0,1)$,
$$f\sim g\Rightarrow \alpha f+(1-\alpha)g\succsim f.$$
\end{description}
In stage 1, the leader takes into account the follower's response and, as the uncertainty seeking self, maximizes $\min_{p\in \Delta}\left(\int u(f)\,dp+c(p)\right)$ over $c\in C$. As a consequence, the Stackelberg equilibrium value of the leader-follower game is \eqref{eq:vf:nive:p} and represents the choice criterion of the decision maker.

Representation \eqref{eq:vf:nive:q} can be similarly interpreted. Now the uncertainty averse self is the leader and the uncertainty seeking self is the follower.
The game has two stages.
\begin{description}
\item[Stage 1:] The leader moves first and chooses a strategy $b\in B$.
\item[Stage 2:] Then in response, the follower (the uncertainty seeking self) moves and chooses a strategy $q\in\Delta$ to minimize $\left(\int u(f)\,dq-b(q)\right)$, acting according to the uncertainty seeking variational preference with penalty function $b$. Here an uncertainty seeking variational preference refers to a preference that satisfies Axioms A1--A4 and the next one:
\item[Axiom A6 (Uncertainty Seeking)] For all $f,g\in\Fc$ and $\alpha\in(0,1)$,
$$f\sim g\Rightarrow f\succsim\alpha f+(1-\alpha)g.$$
\end{description}
In stage 1, the leader takes into account the follower's response  and, as the uncertainty averse self, minimizes $\max_{q\in \Delta}\left(\int u(f)\,dq-b(q)\right)$ over $b\in B$. As a consequence, the Stackelberg equilibrium value of the leader-follower game is \eqref{eq:vf:nive:q} and represents the choice criterion of the decision maker.

As discussed above, each of the representations \eqref{eq:vf:nive:p} and \eqref{eq:vf:nive:q} characterizes explicitly the roles of the uncertainty averse and uncertainty seeking selves in making decisions. The calculus process of each of the representations can be interpreted as a leader-follower game between the two selves.

It is also interesting to relate the representations \eqref{eq:vf:nive:p} and \eqref{eq:vf:nive:q} to zero-sum games. In general, however, both of $C$ and $B$ are nonconvex and hence the values of the zero-sum games do not exist. 
If, for every $f$, the value of the zero-sum game in \eqref{eq:vf:nive:p} exists, that is,
$$\max_{c\in C}\min_{p\in \Delta}\left(\int u(f)dp+c(p)\right)=\min_{p\in \Delta}\sup_{c\in C}\left(\int u(f)dp+c(p)\right),$$
then
\begin{align*}
V(f)= \min_{p\in \Delta}\left(\int u(f)dp+c_0(p)\right),
\end{align*}
where $c_0(p)=\sup_{c\in C}c(p)$ for all $p\in\Delta$. Obviously, $c_0\in\Csc$. 
In this case, by the main representation theorem of MMR (2006), $\succsim$ is an uncertainty averse variational preference.
Similarly,  if the value of the zero-sum game in \eqref{eq:vf:nive:q} exists for every $f$, then $\succsim$ is an uncertainty seeking variational preference. 
Furthermore, if the values of both zero-sum games exist for every $f$, then $\succsim$ is an uncertainty aversion and uncertainty seeking variational preference. Therefore, $\succsim$ satisfies both of the Uncertainty Aversion and Uncertainty Seeking Axioms. As a consequence, under the Weak C-Independence axiom, $\succsim$ satisfies the Independence Axiom of Anscombe and Aumann (1963) and hence is an SEU preference.

\section{Invariant Biseperable Preferences}\label{sec:ibp}


The next axioms was introduced by Gilboa and Schmeidler (1989) in their
characterization of maxmin EU preferences.

\begin{description}
\item[Axiom A7 (C-Independence)] \textit{For all $f,g\in\Fc$, $x\in
X$, and $\alpha\in(0,1)$,
$$f\succsim  g\Leftrightarrow
\alpha f+(1-\alpha)x\succsim\alpha g+(1-\alpha)x.$$}
\end{description}

Following GMM (2004), a binary
relation $\succsim$ is called an \textit{invariant biseparable
preference} (IB preference) if it satisfies Axioms A1--A3 and A7. For discussions on more general
biseparable preferences, see Ghirardato and Marinacci (2001).
Obviously, Axiom A7 implies Axiom A4. Therefore, an IB preference is a special niveloidal preference.
The next theorem characterizes the IB preferences.

\begin{theorem}\label{thm:ibp:rep}
Let $\succsim$ be a binary relation on $\Fc$. The following conditions are equivalent:

\begin{description}
  \item[(i)] $\succsim$ is an IB preference;

  \item[(ii)] There exists a nonconstant affine function $u: X\rightarrow\Rbb$ and a family $\Psc$ consisting of some convex and compact subsets $P\subseteq\Delta$ such that, for any $f$ and $g$ in $\Fc$,
\begin{align}\label{ibp:eq-fg-maxmin}
f\succsim g\Leftrightarrow
\max_{P\in\Psc}\min_{p\in P}\int u(f)\,dp
\ge\max_{P\in\Psc}\min_{p\in P}\int u(g)\,dp;
\end{align}

\item[(iii)] There exists a nonconstant affine function $u: X\rightarrow\Rbb$ and a family $\Qsc$ consisting of some convex and compact subsets $Q\subseteq\Delta$ such that, for any $f$ and $g$ in $\Fc$,
\begin{align}\label{ibp:eq-fg-minmax}
f\succsim g\Leftrightarrow
\min_{Q\in\Qsc}\max_{q\in Q}\int u(f)\,dq
\ge\min_{Q\in\Qsc}\max_{q\in Q}\int u(g)\,dq.
\end{align}
\end{description}

The function $u$ is cardinally unique, there is a
(unique) maximal family $\Psc^*$ satisfying
\eqref{ibp:eq-fg-maxmin}, given by
\begin{align}\label{eq:psc*}
\Psc^*=\left\{P\subseteq\Delta\,\left|\, P\mbox{ is convex and compact, }\min_{p\in P}\int u(f)\,dp\le u(x_f)\mbox{ for all }f\in\Fc\right.\right\},
\end{align}
and there is a
(unique) maximal family $\Qsc^*$ satisfying
\eqref{ibp:eq-fg-minmax}, given by
\begin{align}\label{eq:qsc*}
\Qsc^*=\left\{Q\subseteq\Delta\,\left|\, Q\mbox{ is convex and compact, }\max_{q\in Q}\int u(f)\,dq\ge u(x_f)\mbox{ for all }f\in\Fc\right.\right\}.
\end{align}
\end{theorem}

\proof See Appendix \ref{sec:proofs}. \qed

\medskip

A decision maker with an IB preference ranks acts $f$ according to the preference functional
\begin{align}\label{eq:vf:ibp:p}
V(f)=\max_{P\in\Psc}\min_{p\in P}\int u(f)\,dp
\end{align}
or
\begin{align}\label{eq:vf:ibp:q}
V(f)=\min_{Q\in\Qsc}\max_{q\in Q}\int u(f)\,dq.
\end{align}
Just like the discussion in Section \ref{sec:nive}, each of representations \eqref{eq:vf:ibp:p}--\eqref{eq:vf:ibp:q} can be regarded as the value of a leader-follower game between the uncertainty averse self and the uncertainty seeking self.  For example, the game corresponding to representation \eqref{eq:vf:ibp:p} has two stages.
\begin{description}
\item[Stage 1:]  The leader (the uncertainty seeking self) moves first and chooses a strategy $P\in\Psc$, which is the strategy space of the follower.
\item[Stage 2:] Then in response, the follower (the uncertainty averse self) moves and chooses a strategy $p\in P$ to minimize the expected utility $\int u(f)\,dp$, acting according to maxmin EU with prior set $P$.
\end{description}
In stage 1, the leader takes into account the follower's response and, as the uncertainty seeking self, maximizes $\min_{p\in P}\int u(f)\,dp$ over $P\in\Psc$. As a consequence, the Stackelberg equilibrium value of the leader-follower game is \eqref{eq:vf:ibp:p} and  represents the choice criterion of the decision maker.
Representation \eqref{eq:vf:ibp:q} can be similarly interpreted.

An IB preference has representations given by Theorem \ref{thm:ibp:rep} and, as a special niveloidal preference, it also has representations given by Theorem \ref{thm:nive:rep}. A natural question arises: what is the relationship between these representations?  It can be answered by the next proposition. To proceed, for any subset $P\subseteq\Delta$, let function $\delta_P: \Delta\to(-\infty,\infty]$ be the indicator function of $P$ in the convex analysis, i.e., 
$$\delta_P(p)=\begin{cases}
      0& \text{ if }p\in P, \\
     \infty & \text{otherwise}.
\end{cases}
$$

\begin{proposition}\label{prop:ibp:nive}
Let $\succsim$ be an IB preference on $\Fc$. Let $\Psc^*$, $\Qsc^*$, $C^*$ and $B^*$ are given by Theorems \ref{thm:ibp:rep} and \ref{thm:nive:rep}.
Then we have
\begin{align}\label{eq:C*:P*}
&C^*=\{c\in\Csc\,|\, c\le \delta_P\mbox{ for some }P\in\Psc^*\},\\
&B^*=\{b\in\Csc\,|\, b\ge \delta_Q\mbox{ for some }Q\in\Qsc^*\}.\label{eq:B*:Q*}
\end{align}
\end{proposition}

\proof See Appendix \ref{sec:proofs}. \qed

\medskip

In terms of the indicator functions, \eqref{eq:vf:ibp:p} and \eqref{eq:vf:ibp:q} can be rewritten as
\begin{align}\label{eq:vf:ibp:p1}
V(f)=\max_{P\in\Psc}\min_{p\in \Delta}\left(\int u(f)\,dp+\delta_P(p)\right)
\end{align}
or
\begin{align}\label{eq:vf:ibp:q1}
V(f)=\min_{Q\in\Qsc}\max_{q\in \Delta}\left(\int u(f)\,dq+\delta_Q(q)\right).
\end{align}
They are also related to to zero-sum games,  the values of which, in general, do not exist. 
If, for every $f\in\Fc$, the value of the zero-sum game in \eqref{eq:vf:ibp:p1} exists, that is,
\begin{align*}
\max_{P\in \Psc}\min_{p\in \Delta}\left(\int u(f)dp+\delta_P(p)\right)
= \min_{p\in \Delta}\sup_{P\in \Psc}\left(\int u(f)dp+\delta_P(p)\right),
\end{align*}
then
\begin{align*}
V(f)=\min_{p\in P_0}\int u(f)dp,
\end{align*}
where $P_0=\cap_{P\in\Psc}P$.  
In this case,  $\succsim$ is a maxmin EU preference of Gilboa and Schmeidler (1989).
Similarly,  if the value of the zero-sum game in \eqref{eq:vf:ibp:q1} exists for every $f$, then $\succsim$ is a maxmax EU preference. Furthermore, if the values of both zero-sum games exist for every $f$, then $\succsim$ is an SEU preference.

\section{Special Cases}\label{sec:special}

In this section, we provide explicit characterizations of $\Csc^*$ and $\Bsc^*$ for some special preferences.

\subsection{Variational Preferences}

Let $\succsim$ be an uncertainty averse variational preference of MMR (2006) represented by
\begin{align}\label{eq:int:vp}
\min_{p\in\Delta}\left(\int u(f)\,dp+c_0(p)\right),
\end{align}
where $c_0\in\Csc$ is grounded. 
Now we are going to provide explicit characterizations of $\Csc^*$ and $\Bsc^*$. To this end, for $c_1,c_2\in\Csc$, we write $c_1\apru c_2$ if 
$$\min_{p\in\Delta}\left(\int \varphi\,dp+c_1(p)\right)=\min_{p\in\Delta}\left(\int \varphi\,dp+c_2(p)\right)\quad\mbox{ for all }\varphi\in B_0(\Sigma, u(X)).$$

\begin{proposition}\label{prop:vp:cb}
Consider an uncertainty averse variational preference $\succsim$  of MMR (2006) represented by \eqref{eq:int:vp},
where $c_0\in\Csc$ is grounded. 
Then
\begin{align}\label{eq:C*:vpa}
C^*&=\left\{c\in\Csc\,\left|\, c\apru c_1\le c_0 \mbox{ for some } c_1\in\Csc\right.\right\},\\
B^*&=\left\{b\in\Csc\,\left|\,\min_{p\in\Delta}[b(p)+\hat c(p)]\le0\mbox{ for some }\hat c\in\Csc\mbox{ with } \hat c\apru c_0\right.\right\}.\label{eq:B*:vpa}
\end{align}
\end{proposition}

\proof See Appendix \ref{sec:proofs}.\qed

\begin{remark} If $u(X)$ is unbounded, then $c_1\apru c_2\Leftrightarrow c_1=c_2$ and, therefore, \eqref{eq:C*:vpa}--\eqref{eq:B*:vpa} reduce to
\begin{align*}
C^*&=\left\{c\in\Csc\,\left|\, c\le c_0\right.\right\},\\
B^*&=\left\{b\in\Csc\,\left|\,\min_{p\in\Delta}[b(p)+ c_0(p)]\le0\right.\right\}.\end{align*}
\end{remark} 

Similar to Proposition \ref{prop:vp:cb}, we have the following proposition.

\begin{proposition}
Consider an uncertainty seeking variational preference $\succsim$ that satisfies Axioms A1--A4 and A6 and that is represented by
$$\max_{p\in\Delta}\left(\int u(f)\,dp-b_0(p)\right),$$
where $b_0\in\Csc$ is grounded. For $b_1,b_2\in\Csc$, we write $b_1\uapr b_2$ if 
$$\max_{p\in\Delta}\left(\int \varphi\,dp-b_1(p)\right)=\max_{p\in\Delta}\left(\int \varphi\,dp-b_2(p)\right)\quad\mbox{ for all }\varphi\in B_0(\Sigma, u(X)).$$
Then
\begin{align}\label{eq:C*:vpb}
C^*&=\left\{c\in\Csc\,\left|\, 
\min_{p\in\Delta}[\hat c(p)+b_0(p)]\le0\mbox{ for some }\hat c\in\Csc\mbox{ with } \hat c\apru c\right.\right\},
\\
B^*&=\left\{b\in\Csc\,\left|\,b\le \hat b\uapr b_0\mbox{ for some } \hat b\in\Csc\right.\right\}.\label{eq:B*:vpb}
\end{align}
\end{proposition}

\begin{remark}
If $u(X)$ is unbounded, then $b_1\uapr b_2\Leftrightarrow b_1=b_2$ and, therefore, \eqref{eq:C*:vpb}--\eqref{eq:B*:vpb} reduce to
\begin{align*}
C^*&=\left\{c\in\Csc\,\left|\, 
\min_{p\in\Delta}[c(p)+b_0(p)]\le0\right.\right\},
\\
B^*&=\left\{b\in\Csc\,\left|\,b\le b_0\right.\right\}.
\end{align*}
\end{remark}

\subsection{$\alpha$-Maxmin Expected Utility Preferences}

Consider a preference $\succsim$ that is represented by $\alpha$-maxmin expected utility: 
$$\alpha\min_{p\in P_1}\int u(f)\,dp+(1-\alpha)\max_{p\in P_2}\int u(f)\,dp,\quad  f\in \Fc,$$
where $P_1,P_2\subseteq\Delta$ are convex and compact and $\alpha\in[0,1]$. 
Now $\succsim$ is an IB preference. Therefore, it suffices to characterize $\Psc^*$ and $\Qsc^*$, by Proposition \ref{prop:ibp:nive}.      
We first characterize $\Psc^*$. Assume $P\subseteq\Delta$ is convex and compact. We have
\begin{align*}
&P\in \Psc^*\\
\Leftrightarrow&\min_{p\in P}\int\varphi\,dp\le\alpha\min_{p\in P_1}\int\varphi\,dp+(1-\alpha)\max_{p\in P_2}\int\varphi\,dp\quad \mbox{ for all } \varphi\in B_0(\Sigma)\\
\Leftrightarrow&\alpha\min_{p_1\in P_1}\int\varphi\,dp_1+(1-\alpha)\max_{p_2\in P_2}\int\varphi\,dp_2-\min_{p\in P}\int\varphi\,dp\ge0\quad \mbox{ for all } \varphi\in B_0(\Sigma)\\
\Leftrightarrow&\max_{(p_2,p)\in P_2\times P}\left(\alpha\int\varphi\,dp_1+(1-\alpha)\int\varphi\,dp_2-\int\varphi\,dp\right)\ge0\quad \mbox{ for all } \varphi\in B_0(\Sigma),\, p_1\in P_1\\
\Leftrightarrow&\inf_{\varphi\in B_0(\Sigma)}\max_{(p_2,p)\in P_2\times P}\left(\alpha\int\varphi\,dp_1+(1-\alpha)\int\varphi\,dp_2-\int\varphi\,dp\right)\ge0,\quad \forall\, p_1\in P_1\\
\Leftrightarrow&\max_{(p_2,p)\in P_2\times P}\inf_{\varphi\in B_0(\Sigma)}\left(\alpha\int\varphi\,dp_1+(1-\alpha)\int\varphi\,dp_2-\int\varphi\,dp\right)\ge0,\quad \forall\, p_1\in P_1,
\end{align*}
where the minimax theorem is used in the last equivalence. 
If $\alpha p_1+(1-\alpha)p_2\ne p$, then
$$\inf_{\varphi\in B_0(\Sigma)}\left(\alpha\int\varphi\,dp_1+(1-\alpha)\int\varphi\,dp_2-\int\varphi\,dp\right)=-\infty.$$
Moreover, if $\alpha p_1+(1-\alpha)p_2=p$, then
$$\alpha\int\varphi\,dp_1+(1-\alpha)\int\varphi\,dp_2-\int\varphi\,dp=0.$$
Therefore,
\begin{align*}
P\in \Psc^*\Leftrightarrow &\ \forall\,p_1\in P_1,\,\exists\, (p_2,p)\in P_2\times P \mbox{ s.t. }p=\alpha p_1+(1-\alpha)p_2,\\
\Leftrightarrow &\ \alpha P_1\subseteq P-(1-\alpha)P_2. 
\end{align*}
That is,
$$\Psc^*=\{P\subseteq\Delta\,|\,P\mbox{ is convex and compact, }\alpha P_1\subseteq P-(1-\alpha)P_2\}.$$
Similarly, we have
$$\Qsc^*=\{Q\subseteq\Delta\,|\,Q\mbox{ is convex and compact, }(1-\alpha) P_2\subseteq Q-\alpha P_1\}.$$
Particularly, if $P_1=P_2=P_0$, then 
\begin{align*}
&\Psc^*=\{P\subseteq\Delta\,|\,P\mbox{ is convex and compact, }\alpha P_0\subseteq P-(1-\alpha)P_0\},\\
&\Qsc^*=\{Q\subseteq\Delta\,|\,Q\mbox{ is convex and compact, }(1-\alpha) P_0\subseteq Q-\alpha P_0\}.
\end{align*}
In the case of $\alpha=1$, the preference is the maxmin EU of Gilboa and Schmeidler (1989) and
\begin{align*}
\Psc^*&=\{P\subseteq\Delta\,|\, P\mbox{ is convex and compact, } P_0\subseteq P\},\\
\Qsc^*&=\{Q\subseteq\Delta\,|\, Q\mbox{ is convex and compact, } Q\cap P_0\neq\emptyset\}.
\end{align*}
In the case of $\alpha=0$, the preference is the maxmax EU and 
\begin{align*}
\Psc^*&=\{P\subseteq\Delta\,|\, P\mbox{ is convex and compact, } P\cap P_0\neq\emptyset\},\\
\Qsc^*&=\{Q\subseteq\Delta\,|\, Q\mbox{ is convex and compact, } P_0\subseteq Q\}.
\end{align*}
Moreover, for an SEU preference $\succsim$ represented by $\int u(f)\,dp_0$ with $p_0\in\Delta$, we have
\begin{align}\label{eq:seup0}
\Psc^*=\Qsc^*=\{P\subseteq\Delta\,|\, P\mbox{ is convex and compact, } p_0\in P\}.
\end{align}

\subsection{Choquet Expected Utility Preferences}

Consider a preference $\succsim$ that is represented by Choquet expected utility of Schmeidler (1989): 
$$\int u(f)\,d\pi,\quad f\in \Fc,$$
where $\pi$ is a capacity on $(S,\Sigma)$ and $\int u(f)d\pi$ is the Choquet expectation of $u(f)$ w.r.t. $\pi$. 
Now $\succsim$ is an IB preference. Therefore, it suffices to characterize $\Psc^*$ and $\Qsc^*$, by Proposition \ref{prop:ibp:nive}.  
We first characterize $\Psc^*$. Assume $P\subseteq\Delta$ is convex and compact. We have
\begin{align*}
&P\in \Psc^*\\
\Leftrightarrow&\min_{p\in P}\int\varphi\,dp\le\int\varphi\,d\pi\quad \mbox{for all } \varphi\in B_0(\Sigma)\\
\Leftrightarrow&\int\varphi\,d\pi-\min_{p\in P}\int\varphi\,dp\ge0\quad \mbox{for all } \varphi\in B_0^+(\Sigma)\\
\Leftrightarrow&\max_{p\in P}\left(\int\varphi\,d\pi-\int\varphi\,dp\right)\ge0\quad \mbox{for all } \varphi\in B_0^+(\Sigma)\\
\Leftrightarrow&\inf_{\varphi\in [\phi]}\max_{p\in P}\left(\int\varphi\,d\pi-\int\varphi\,dp\right)\ge0\quad\mbox{for all }  \phi\in B_0^+(\Sigma)\\
\Leftrightarrow&\max_{p\in P}\inf_{\varphi\in [\phi]}\left(\int\varphi\,d\pi-\int\varphi\,dp\right)\ge0\quad \mbox{for all }  \phi\in B_0^+(\Sigma),
\end{align*}
where the minimax theorem is used in the last equivalence and $[\phi]$ denotes the convex set given by
$$[\phi]\trieq\left\{\varphi\in B_0^+(\Sigma)\,\left|\, 
\varphi=k(\phi)\mbox{ for some non-decreasing function }k:\Rbb^+\to\Rbb^+\right.\right\}.$$
A \textit{finite chain} in $\Sigma$ is a finite sequence $\{A_i, 0\le i\le n\}\subset\Sigma$ that satisfies $$\emptyset=A_0\subset A_1\subset A_2\subset\cdots\subset A_n=S.$$ 
For every $\phi\in B_0^+(\Sigma)$, there exist a finite chain $\{A_i, 0\le i\le n\}$ and a decreasing sequence $\alpha_1>\alpha_2>\cdots>\alpha_n\ge0$ such that
$$\phi(s)=\alpha_i\quad\mbox{ for }s\in A_i\backslash A_{i-1}, 1\le i\le n.$$
In this case, we have $\varphi\in[\phi]$ if and only if there exists a non-increasing sequence $\beta_1\ge \beta_2\ge\cdots\ge\beta_n\ge0$ such that
$$\varphi(s)=\beta_i\quad\mbox{ for }s\in A_i\backslash A_{i-1}, 1\le i\le n.$$
Moreover,
$$\int\varphi\,d\pi=\sum_{i=1}^n(\beta_i-\beta_{i+1})\pi(A_i),$$
where $\beta_{n+1}\trieq0$. Therefore, assume $P\subseteq\Delta$ is convex and compact, then
\begin{align*}
&P\in \Psc^*\\
\Leftrightarrow&\,\forall\,\mbox{finite chain }\{A_i, 0\le i\le n\}, \exists\, p\in P\mbox{ s.t. }\sum_{i=1}^n(\beta_i-\beta_{i+1})[\pi(A_i)-p(A_i)]\ge0\\
&\mbox{ for all non-increasing sequence }\beta_1\ge \beta_2\ge\cdots\ge\beta_n\ge\beta_{n+1}=0\\
\Leftrightarrow&\,\forall\,\mbox{finite chain }\{A_i, 0\le i\le n\}, \exists\, p\in P\mbox{ s.t. }p(A_i)\le\pi(A_i)\mbox{ for all }1\le i\le n\\
\Leftrightarrow&\,P\bigcap\{p\in\Delta\,|\,p(A_i)\le\pi(A_i),\ 1\le i\le n\}\neq\emptyset \mbox{ for all finite chain }\{A_i, 0\le i\le n\}.
\end{align*}
Similarly, assume $Q\subseteq\Delta$ is convex and compact, then \begin{align*}
&Q\in \Qsc^*\\
\Leftrightarrow&\,Q\bigcap\{q\in\Delta\,|\,q(A_i)\ge\pi(A_i),\ 1\le i\le n\}\neq\emptyset \mbox{ for all finite chain }\{A_i, 0\le i\le n\}.
\end{align*}

\section{Ambiguity Attitude}\label{sec:amg}

In this section we discuss the attitude towards ambiguity of
decision makers with niveloidal preferences, following the approach
proposed in Ghirardato and Marinacci (2002).
First the comparative notion of ambiguity attitude is stated as
follows:

\begin{definition}\label{def:more:averse}
Let $\succsim_1$ and $\succsim_2$ be preference relations on $\Fc$.
We say $\succsim_1$ is \textit{more ambiguity averse} than
$\succsim_2$ if, for all $f\in\Fc$ and $x\in X$,
$$x\succsim_2 f\Rightarrow x\succsim_1 f.$$
\end{definition}

For simplicity, we write $u_1\approx u_2$ to denote that utility indices $u_1$
and $u_2$ are cardinally equivalent, i.e., they are positive affine
transformations of each other. The proofs of the next three propositions
are standard\footnote{Following the same procedure as in, e.g., MMR (2006); see also C3M (2011).} and omitted.

\begin{proposition}\label{prop:comp:nive}
Let $\succsim_1$ and $\succsim_2$ be two niveloidal preferences. Let the corresponding utility indices $u_i$ and the  grounded subsets $C_i^*$ and $B_i^*$ of $\Csc$ be given by Theorem \ref{thm:nive:rep}.
Then the following conditions are equivalent:
\begin{description}
  \item[(i)] $\succsim_1$ is more ambiguity averse than $\succsim_2$;
  \item[(ii)] $u_1\approx u_2$ and $C_1^*\subseteq C_2^*$ (provided $u_1=u_2$);
  \item[(iii)] $u_1\approx u_2$ and $B_1^*\supseteq B_2^*$ (provided $u_1=u_2$).
\end{description}
\end{proposition}

\begin{proposition}\label{prop:comp:ibp}
Let $\succsim_1$ and $\succsim_2$ be two IB preferences. Let the corresponding utility indices $u_i$ and the families $\Psc_i^*$ and $\Qsc_i^*$ be given by Theorem \ref{thm:ibp:rep}, $i=1,2$.
Then the following conditions are equivalent:
\begin{description}
  \item[(i)] $\succsim_1$ is more ambiguity averse than $\succsim_2$;
  \item[(ii)] $u_1\approx u_2$ and $\Psc_1^*\subseteq \Psc_2^*$  (provided $u_1=u_2$);
  \item[(iii)] $u_1\approx u_2$ and $\Qsc_1^*\supseteq \Qsc_2^*$  (provided $u_1=u_2$).
\end{description}
\end{proposition}

Proposition \ref{prop:comp:nive} shows that the leader's strategy space in the leader-follower game as discussed as in Section \ref{sec:nive}
serves as an index of the decision maker's ambiguity aversion. More precisely, if the ambiguity seeking self is the leader, then the set $C^*$ serves as an index of the decision maker's ambiguity aversion: the smaller the set $C^*$ is, the more ambiguity averse the decision maker is. Similarly, if the ambiguity averse self is the leader, then the set $B^*$ serves as an index of the decision maker's ambiguity aversion: the larger the set $B^*$ is, the more ambiguity averse the decision maker is. The implications of Proposition \ref{prop:comp:ibp} is similar.

To introduce an absolute notion of ambiguity aversion, we follow
Ghirardato and Marinacci (2002) to consider SEU preferences as
benchmarks for ambiguity neutrality. We then say a preference
$\succsim$ is \textit{ambiguity averse} if it is more ambiguity averse
than some SEU preference.

\begin{proposition}\label{prop:averse:nive}
Let $\succsim$ be a niveloidal preference. Let  $C^*$ and $B^*$  be given by Theorem \ref{thm:nive:rep}. Then the following conditions are equivalent:
\begin{description}
  \item[(i)] $\succsim$ is ambiguity averse;
  \item[(ii)] $\bigcap_{c\in C^*}\{p\in\Delta\,|\, c(p)\le0\}\neq\emptyset$;
  \item[(iii)] $\delta_{\{p_0\}}\in B^*$ for some $p_0\in\Delta$.
\end{description}
\end{proposition}

\proof See Appendix \ref{sec:proofs}.\qed

\medskip

Similarly, we have the following proposition.
\begin{proposition}\label{prop:averse:ibp}
Let $\succsim$ be an IB preference. Let  $\Psc^*$ and $\Qsc^*$ be given by Theorem \ref{thm:ibp:rep}. Then the following conditions are equivalent:
\begin{description}
  \item[(i)] $\succsim$ is ambiguity averse;
  \item[(ii)] $\bigcap_{P\in \Psc^*}P\neq\emptyset$;
  \item[(iii)] $\Qsc^*$ contains a singleton.
\end{description}
\end{proposition}

\newpage

\appendix

\section*{Appendix}

\section{Niveloids}\label{sec:mathpre}

In this appendix, we provide the representations of niveloids.
Let $\Phi$ be a nonempty subset of $B(\Sigma)$. A function
$I: \Phi\rightarrow[-\infty,\infty]$ is called:
\begin{itemize}
  \item \textit{positively homogeneous} if
  $I(\alpha\varphi)=\alpha I(\varphi)$ for all $\varphi\in \Phi$ and $\alpha\ge0$
  with $\alpha\varphi\in\Phi$.

  \item \textit{super-additive} if $I(\varphi+\psi)\ge I(\varphi)+I(\psi)$
  for all $\varphi,\psi\in\Phi$ with $\varphi+\psi\in\Phi$.
\end{itemize}

For the next lemma, see, e.g.,  F\"ollmer and Schied (2016, Propositions 4.6--4.7) and Cerreia-Vioglio, Maccheroni, Marinacci and Rustichini (2014, Proposition 1).

\begin{lemma}\label{lma:niveloid:set}
A functional $I:B(\Sigma)\to\Rbb$ is a niveloid if and only if there exists a nonempty subset $\Phi\subseteq B(\Sigma)$ such that
\begin{itemize}
\item $\sup\{\alpha\in\Rbb\,|\,-\alpha\in\Phi\}<\infty$;
\item For all $\phi\in\Phi$ and $\psi\in B(\Sigma)$, $\psi\ge\phi\Rightarrow \psi\in\Phi$;
\item $I(\phi)=\sup\{\alpha\in\Rbb\,|\,\phi-\alpha\in\Phi\}$.
\end{itemize}
If it is the case, the set $\Phi$ can be chosen as $\Phi_0=\{\phi\in B(\Sigma)\,|\, I(\phi)\ge0\}$. Moreover, we have
\begin{itemize}
\item $I$ is positively homogeneous if and only if $\Phi_0$ is a cone;
\item $I$ is concave if and only if $\Phi_0$ is a convex set;
\item $I$ is positively homogeneous and concave if and only if $\Phi_0$ is a convex cone.
\end{itemize}
\end{lemma}

For any interval $T\subseteq\Rbb$, we use
$\Ic_{cc}(T)$ (resp. $\Ic_{cv}(T)$) to denote all  concave (resp. convex) niveloids $I: B_0(\Sigma, T)\to\Rbb$. The next lemma provides representations of niveloids.

\begin{lemma}\label{lma:I:niveloid}
Let $I: B_0(\Sigma,T)\to\Rbb$ be a functional, where
$T\subseteq\Rbb$ is an interval and $0\in\intr(T)$. Then the
following assertions are equivalent:
\begin{description}
\item[(i)] $I$ is a normalized niveloid;
\item[(ii)] There exists a family $\Jsc\subseteq\Ic_{cc}(T)$ such that
$\sup_{J\in\Jsc}J(0)=0$ and
\begin{equation}\label{eq:IJ:niveloid}
I(\varphi)=\sup_{J\in\Jsc}J(\varphi)\quad\mbox{ for all }\varphi\in B_0(\Sigma,T);
\end{equation}
\item[(iii)] There exists a family $\Ksc\subseteq\Ic_{cv}(T)$ such that
$\inf_{K\in\Ksc}K(0)=0$ and
\begin{equation}\label{eq:IH:niveloid}
I(\varphi)=\inf_{K\in\Ksc}K(\varphi)\quad\mbox{ for all }\varphi\in B_0(\Sigma,T).
\end{equation}
\end{description}
Moreover, the families $\Jsc$ and $\Ksc$ can be chosen such that the supremum in \eqref{eq:IJ:niveloid} and the infimum in \eqref{eq:IH:niveloid} are attainable.
\end{lemma}

\proof Obviously, (ii)$\Rightarrow$(i) and (iii)$\Rightarrow$(i). \begin{description}
\item[(i)$\Rightarrow$(ii):] It proceeds in several steps
as follows. Assume (i) holds.

\item[An extension $\hat I$ of $I$.] Let $\hat I: B(\Sigma)\to\Rbb$
be given as
\begin{align}\label{eq:I:extension}
\hat I(\psi)=\sup_{\varphi\in B_0(\Sigma,
T)}\left[I(\varphi)+\inf_{s\in S}(\psi(s)-\varphi(s))\right], \quad
\psi\in B(\Sigma).\end{align}
Then $\hat I$ is the least niveloid on
$B(\Sigma)$ that extends $I$; see Dolecki and Greco (1995) or
MMR (2006, Appendix A). It is easy
to see that $\hat I$ is a normalized niveloid.

\item[Representation of $\hat I$.]\footnote{The argument of this step was used in Mao and Wang (2020) and Jia, Xia and Zhao (2020) to investigate risk measures.} Let $\Phi^+$ be the upper $0$-level set of $\hat I$, i.e., 
$$\Phi^+=\left\{\varphi\in B(\Sigma)\ |\ \hat I(\varphi)\ge 0\right\}.$$
For every $\varphi\in\Phi^+$, let $\Psi(\varphi)$ be given as
$$\Psi(\varphi)=\left\{\phi\in B(\Sigma)\ |\ \phi\ge\varphi\right\}.$$ Obviously, $\varphi\in\Psi(\varphi)\subseteq\Phi^+$ for every $\phi\in\Phi^+$. Therefore,
$$\Phi^+=\bigcup_{\varphi\in\Phi^+}\{\varphi\}=\bigcup_{\varphi\in\Phi^+}\Psi(\varphi).$$
For every $\varphi\in\Phi^+$, let $J_\varphi:B(\Sigma)\to\Rbb$ be
defined as
$$J_\varphi(\psi)=\sup\left\{\alpha\in\Rbb\ |\ \psi-\alpha\in\Psi(\varphi)\right\}.$$
Then 
$$J_\varphi(\psi)=\inf_{s\in S}(\psi(s)-\varphi(s)),\quad \psi\in B(\Sigma).$$
Obviously, $J_\varphi$ is a concave niveloid. 
For every $\psi\in B(\Sigma)$,
\begin{equation}\label{eq:hatI}
\begin{split}
\hat I(\psi)=&\sup\left\{\alpha\in\Rbb\ \left|\ \hat I(\psi-\alpha)\ge 0\right.\right\}\\
=&\sup\left\{\alpha\in\Rbb\ \left|\
\psi-\alpha\in\Phi^+\right.\right\}\\
=&\sup\left\{\alpha\in\Rbb\ \left|\
\psi-\alpha\in\bigcup_{\varphi\in\Phi^+}\Psi(\varphi)\right.\right\}\\
=&\sup_{\varphi\in\Phi^+}\sup\left\{\alpha\in\Rbb\ |\ \psi-\alpha\in
\Psi(\varphi)\right\}\\
=&\sup_{\varphi\in\Phi^+}J_\varphi(\psi).
\end{split}
\end{equation}

\item[Representation of $I$.] Finally, restricting \eqref{eq:hatI} on $B_0(\Sigma, T)$ leads to
$$I(\psi)= \sup_{\varphi\in\Phi^+}J_\varphi(\psi)\quad \mbox{ for all }\psi\in B_0(\Sigma,
T),$$ where $J_\varphi: B_0(\Sigma, T)\to\Rbb$ is a  concave niveloid for every $\varphi\in\Phi^+$. Obviously, $\sup_{\varphi\in\Phi^+}J_\varphi(0)=I(0)=0$.

\item[(i)$\Rightarrow$(iii):] Assume (i) holds.
Let $J(\varphi)=-I(-\varphi)$ for all $\varphi\in B_0(\Sigma,-T)$. Then
$J: B_0(\Sigma,-T)\to\Rbb$ is a normalized niveloid. Then by
(i)$\Rightarrow$(ii), there exists a family $\{J_\lambda,
\lambda\in\Lambda\}\subseteq\Ic_{cc}(-T)$ such that $\sup_{\lambda\in\Lambda}J_\lambda(0)=0$ and
\begin{equation*}
J(\varphi)=\sup_{\lambda\in\Lambda}J_\lambda(\varphi),\quad
\varphi\in B_0(\Sigma,-T).
\end{equation*}
Let $K_\lambda(\varphi)=-J_\lambda(-\varphi)$ for all $\varphi\in
B_0(\Sigma,T)$, then
$K_\lambda\in\Ic_{cv}(T)$,
$$\inf_{\lambda\in\Lambda}K_\lambda(0)=-\sup_{\lambda\in\Lambda}J_\lambda(0)=0,$$ and
$$I(\varphi)=-J(-\varphi)=-\sup_{\lambda\in\Lambda}J_\lambda(-\varphi)=\inf_{\lambda\in\Lambda}K_\lambda(\varphi).$$
\end{description}
Now we show that the supremum in \eqref{eq:IJ:niveloid} is attainable. Actually, for each $\psi\in B_0(\Sigma,T)$, let $\varphi=\psi-I(\psi)$. Then $\varphi\in\Phi^+$ and
$$J_\varphi(\psi)=\sup\{\alpha\in\Rbb\,|\,\psi-\alpha\ge \varphi\}=I(\psi),$$
which implies that the supremum in \eqref{eq:IJ:niveloid} is attainable. Similarly,
the infimum in \eqref{eq:IH:niveloid} is also attainable.
\qed

\section{Proofs}\label{sec:proofs}

\subsection{Proof of Theorem \ref{thm:nive:rep}}  It is obvious that (ii)$\Rightarrow$(i) and (iii)$\Rightarrow$(i).
\begin{description}
\item[(i)$\Rightarrow$(ii):]  Assume (i) holds. By Lemma \ref{lma:niveloid}, there
exists a nonconstant affine function $u: X\rightarrow\Rbb$ and a normalized niveloid $I: B_0(\Sigma, u(X))\to\Rbb$ such that
$0\in\mathrm{int}(u(X))$ and, for all
$f,g\in\Fc$,
$$f\succsim g\ \Leftrightarrow\ I(u(f))\ge I(u(g)).$$
By Lemma \ref{lma:I:niveloid}, there exists a family $\Jsc\subseteq\Ic_{cc}(u(X))$ such that
\begin{align}\label{eq:IJ:ground}
I(\varphi)=\max_{J\in\Jsc}J(\varphi)\quad\mbox{ for all }\varphi\in B_0(\Sigma,u(X)).
\end{align}
By MMR (2006, Lemma 26), 
for every $J\in\Jsc$, there exists a function $c_J\in\Csc$ such that
$$J(\varphi)=\min_{p\in\Delta}\left(\int\varphi\, dp+c_J(p)\right)\quad \mbox{ for all }\varphi\in B_0(\Sigma,u(X)).$$
Therefore,  by \eqref{eq:IJ:ground}, we have
$$I(\varphi)=\max_{J\in\Jsc}\min_{p\in\Delta}\left(\int\varphi\, dp+c_J(p)\right)\quad \mbox{ for all }\varphi\in B_0(\Sigma,u(X))$$
and $$\max_{J\in\Jsc}\min_{p\in\Delta}c_J(p)=I(0)=0,$$
which implies (ii).
\item[(i)$\Rightarrow$(iii):] It is similar to (i)$\Rightarrow$(ii).
\end{description}


Let $C^*$ be given by \eqref{eq:c*}. Now we show that $C^*$ is the maximal grounded subset of $\Csc$ satisfying \eqref{nive:eq-fg-maxmin}. Actually,
let  $C$ be a grounded subset of $\Csc$ such that \eqref{nive:eq-fg-maxmin} holds for any $f$ and $g$ in $\Fc$. Then, for all $f\in\Fc$, 
\begin{align*}
\max_{c\in C}\min_{p\in \Delta}\left(\int u(f)\,dp+c(p)\right)=\max_{c\in C}\min_{p\in \Delta}\left(\int u(x_f)\,dp+c(p)\right)=u(x_f),
\end{align*}
where $x_f$ is the certainty equivalent of $f$. Then by \eqref{eq:c*}, we have $C\subseteq C^*$. Moreover,
\begin{align*}
u(x_f)&\ge\sup_{c\in C^*}\min_{p\in \Delta}\left(\int u(f)\,dp+c(p)\right)\\
&\ge \max_{c\in C}\min_{p\in \Delta}\left(\int u(f)\,dp+c(p)\right)\\
&=u(x_f)\quad\mbox{ for all } f\in\Fc,\end{align*}
which implies that
$$\max_{c\in C^*}\min_{p\in \Delta}\left(\int u(f)\,dp+c(p)\right)=u(x_f)=I(u(x_f))=I(u(f))\quad\mbox{ for all } f\in\Fc.$$
Therefore, $C^*$ satisfies \eqref{nive:eq-fg-maxmin}.

By a similar way, we can show that $B^*$, which is given by \eqref{eq:b*}, is the maximal grounded subset of $\Csc$ satisfying \eqref{nive:eq-fg-minmax}. \qed

\subsection{Proofs of Theorem \ref{thm:ibp:rep} and Proposition \ref{prop:ibp:nive}}

The following representation result can be easily proved by
mimicking the arguments of Gilboa and Schmeidler (1989, Lemmas
3.1--3.3), see also GMM (2004, Lemma 1).

\begin{lemma}\label{lma:Iu:clinear}
A binary relation $\succsim$ on $\Fc$ is an IB preference if and
only if there exists a nonconstant affine function $u: X\to\Rbb$ and a positively homogeneous niveloid $I: B_0(\Sigma)\to\Rbb$
such that, for any $f,g\in\Fc$,
\begin{equation*}
f\succsim g\Leftrightarrow I(u(f))\ge I(u(g)).
\end{equation*}
Moreover,  $u$ is cardinally unique 
and, given $u$, $I$ is unique.
\end{lemma}

\paragraph{Proof of Theorem \ref{thm:ibp:rep}.}  It is obvious that (ii)$\Rightarrow$(i) and (iii)$\Rightarrow$(i).
\begin{description}
\item[(i)$\Rightarrow$(ii):]  Assume (i) holds. By Lemma \ref{lma:Iu:clinear}, there
exists a nonconstant affine function $u: X\rightarrow\Rbb$ and a positively homogeneous niveloid $I: B_0(\Sigma)\rightarrow\Rbb$ such that $0\in\mathrm{int}(u(X))$ and, for all
$f,g\in\Fc$,
$$f\succsim g\ \Leftrightarrow\ I(u(f))\ge I(u(g)).$$

Let $C^*$ be given by Theorem \ref{thm:nive:rep}.
Then the uniqueness of $I$ implies that
\begin{align*}
I(\phi)=\max_{c\in C^*}\min_{p\in \Delta}\left(\int \phi\, dp+c(p)\right)\quad\mbox{ for all }\phi\in B_0(\Sigma, u(X)).
\end{align*}
For every $c\in C^*$, let
$$I_c(\phi)=\min_{p\in \Delta}\left(\int \phi\, dp+c(p)\right)\quad\mbox{ for all }\phi\in B_0(\Sigma, u(X)).$$
Obviously, $I_c$ is a concave niveloid on $B_0(\Sigma,u(X))$ and $I_c\le I$  on $B_0(\Sigma,u(X))$.
We can extend $I$ (resp. $I_c$) from $B_0(\Sigma,u(X))$ to $B(\Sigma)$ by the same way as in \eqref{eq:I:extension} and denote the extension by $\hat I$ (reap. $\hat I_c$). We can see that $\hat I$ is a positively homogeneous niveloid on $B(\Sigma)$ and $\hat I_c$ is a concave niveloid on $B(\Sigma)$. Moreover, $\hat I_c\le \hat I$.
Let
\begin{align*}
\Phi_c&=\{\phi\in B(\Sigma)\,|\, \hat I_c(\phi)\ge 0\},\\
\Phi&=\{\phi\in B(\Sigma)\,|\, \hat I(\phi)\ge 0\}.
\end{align*}
Then Lemma \ref{lma:niveloid:set}
 implies that $\Phi_c$ is convex, $\Phi$ is a cone, $\Phi_c\subseteq\Phi$, and
\begin{align*}
\hat I_c(\phi)&=\sup\{\alpha\in\Rbb\,|\,\phi-\alpha\in\Phi_c\}\quad\mbox{ for all }\phi\in B(\Sigma),\\
\hat I(\phi)&=\sup\{\alpha\in\Rbb\,|\,\phi-\alpha\in\Phi\}\quad\mbox{ for all }\phi\in B(\Sigma).
\end{align*}
Let $\cone(\Phi_c)$ denote the cone generated by $\Phi_c$. Then $\Phi_c\subseteq\cone(\Phi_c)\subseteq\Phi$ and  $\cone(\Phi_c)$ is a convex cone. Let
$$J(\phi)=\sup\{\alpha\in\Rbb\,|\,\phi-\alpha\in\cone(\Phi_c)\}\quad\mbox{ for all }\phi\in B(\Sigma).$$
Then Lemma \ref{lma:niveloid:set}
 implies that $J$ is a positively homogeneous and concave niveloid on $B(\Sigma)$ and $\hat I_c\le J\le\hat I$. By Gilboa and Schmeidler (1989, Lemma 3.5), there exists a unique convex and compact subset $P\subseteq\Delta$ such that
$$J(\phi)=\min_{p\in P}\int\phi\, dp\quad\mbox{ for all }\phi\in B(\Sigma).$$
Obviously,
$$\min_{p\in \Delta}\left(\int \phi\, dp+c(p)\right)=I_c(\phi)=\hat I_c(\phi)\le J(\phi)=\min_{p\in P}\int\phi\, dp\le \hat I(\phi)=I(\phi)$$
for every $\phi\in B_0(\Sigma,u(X))$. 
Therefore,
\begin{align*}
I(\phi)=\max_{c\in C^*}\min_{p\in \Delta}\left(\int \phi\, dp+c(p)\right)=\max_{P\in\Psc}\min_{p\in P}\int\phi\, dp\quad\mbox{ for all }\phi\in B_0(\Sigma, u(X)).
\end{align*}
This concludes (i)$\Rightarrow$(ii).

\item[(i)$\Rightarrow$(iii):] It is similar to (i)$\Rightarrow$(ii).
\end{description}


Let $\Psc^*$ be given by \eqref{eq:psc*}. Now we show that $\Psc^*$ is the maximal family satisfying \eqref{ibp:eq-fg-maxmin}. Actually,
let  $\Psc$ be a family consisting of some convex and compact subsets $P\subseteq\Delta$ such that \eqref{ibp:eq-fg-maxmin} holds for any $f$ and $g$ in $\Fc$. Then
\begin{align*}
\max_{P\in\Psc}\min_{p\in P}\int u(f)\,dp=\max_{P\in\Psc}\min_{p\in P}\int u(x_f)\,dp=u(x_f),\ \forall\,f\in\Fc,
\end{align*}
where $x_f$ is the certainty equivalent of $f$. Then by \eqref{eq:psc*}, we have $\Psc\subseteq\Psc^*$. Moreover,
$$u(x_f)\ge\sup_{P\in\Psc^*}\min_{p\in P}\int u(f)\,dp\ge \max_{P\in\Psc}\min_{p\in P}\int u(f)\,dp=u(x_f)\quad\mbox{ for all }  f\in\Fc,$$
which implies that
$$\max_{P\in\Psc^*}\min_{p\in P}\int u(f)\,dp=u(x_f)=I(u(x_f))=I(u(f))\quad\mbox{ for all } f\in\Fc.$$
Therefore, $\Psc^*$ satisfies \eqref{ibp:eq-fg-maxmin}.

By a similar way, we can show that $\Qsc^*$, which is given by \eqref{eq:qsc*}, is the maximal family satisfying \eqref{ibp:eq-fg-minmax}. \qed

\paragraph{Proof of Proposition \ref{prop:ibp:nive}.}
We prove \eqref{eq:C*:P*} only, since \eqref{eq:B*:Q*} can be similarly proved. The ``$\supseteq$" part of \eqref{eq:C*:P*} is obvious. It is left to prove the ``$\subseteq$" part. Now assume $c\in C^*$. Let $\hat I_c$, $J$ and $P$ be given as in the proof of (i)$\Rightarrow$(ii) part of Theorem \ref{thm:ibp:rep}. Then $P\in\Psc^*$ and, by MMR (2006, Lemma 26),
$$
c(p)=\sup_{\phi\in B(\Sigma)}\left(\hat I_c(\phi)-\int\phi\, dp\right)
\le\sup_{\phi\in B(\Sigma)}\left(J(\phi)-\int\phi\, dp\right)
\le 0, \quad\forall\,p\in P,
$$ which implies that $c\le \delta_{_P}$.
\qed

\subsection{Proof of Proposition \ref{prop:vp:cb}}  
We first show \eqref{eq:C*:vpa}. 
Assume $c\in\Csc$. We have
\begin{align*}
c\in C^*&\Leftrightarrow \min_{p\in\Delta}\left(\int u(f)\,dp+c(p)\right)\le u(x_f)=\min_{p\in\Delta}\left(\int u(f)\,dp+c_0(p)\right),\ \forall\,f\in\Fc\\
&\Leftrightarrow \min_{p\in\Delta}\left(\int \varphi\,dp+c(p)\right)\le \min_{p\in\Delta}\left(\int \varphi\,dp+c_0(p)\right),\ \forall\,\varphi\in B_0(\Sigma, u(X)).
\end{align*}
Let $I: B_0(\Sigma, u(X))\to\Rbb$ be given by
$$I(\varphi)=\min_{p\in\Delta}\left(\int \varphi\,dp+c(p)\right)\quad\mbox{ for all }\varphi\in B_0(\Sigma, u(X)).$$
Let $\hat I$ be the least niveloid on $B_0(\Sigma)$ that extends $I$. Let $$\hat c(p)=\sup_{\varphi\in B_0(\Sigma)}\left(\hat I(\varphi)-\int\varphi\,dp\right)\quad\mbox{ for all }p\in\Delta.$$
Then $c\apru \hat c$ and
\begin{align*}
c\in C^*\Rightarrow\ &\hat I(\varphi)\le \min_{p\in\Delta}\left(\int \varphi\,dp+c_0(p)\right)\quad\mbox{ for all }\varphi\in B_0(\Sigma)\\
\Rightarrow\ &\hat c\le c_0,
\end{align*} 
which implies the ``$\subseteq$" part of \eqref{eq:C*:vpa}.
Conversely, the ``$\supseteq$" part of \eqref{eq:C*:vpa} is obvious. 
Therefore, \eqref{eq:C*:vpa} is proved.

Next we show \eqref{eq:B*:vpa}. Assume $b,c\in\Csc$. By the minimax theorem,\footnote{See, e.g., Mertens, Sorin and Zamir (2015, Theorem I.1.1).} we have
\begin{align*}
&\inf_{\varphi\in B_0(\Sigma)}\left\{\max_{p\in\Delta}\left(\int\varphi\,dp-b(p)\right)-\min_{q\in\Delta}\left(\int\varphi\,dq+c(q)\right)\right\}\\
=&\inf_{\varphi\in B_0(\Sigma)}\max_{(p,q)\in\Delta^2}\left(\int\varphi\,dp-\int\varphi\,dq-b(p)-c(q)\right)\\
=&\max_{(p,q)\in\Delta^2}\inf_{\varphi\in B_0(\Sigma)}\left(\int\varphi\,dp-\int\varphi\,dq-b(p)-c(q)\right).
\end{align*}
If $p,q\in\Delta$ and $p\ne q$, then 
$$\inf_{\varphi\in B_0(\Sigma)}\left(\int\varphi\,dp-\int\varphi\,dq\right)=-\infty,$$
which implies
\begin{align}\label{eq:fen:dual}
\begin{split}
&\inf_{\varphi\in B_0(\Sigma)}\left\{\max_{p\in\Delta}\left(\int\varphi\,dp-b(p)\right)-\min_{q\in\Delta}\left(\int\varphi\,dq+c(q)\right)\right\}\\
=&\max_{p=q\in\Delta}\inf_{\varphi\in B_0(\Sigma)}\left(\int\varphi\,dp-\int\varphi\,dq-b(p)-c(q)\right)\\
=&-\min_{p\in\Delta}[(b(p)+c(p)].
\end{split}
\end{align}
Let $I_0: B_0(\Sigma, u(X))\to\Rbb$ be given by
$$I_0(\varphi)=\min_{p\in\Delta}\left(\int \varphi\,dp+c_0(p)\right)\quad\mbox{ for all }\varphi\in B_0(\Sigma, u(X)).$$
Let $\hat I_0$ be the least niveloid on $B_0(\Sigma)$ that extends $I_0$ and
$\hat c_0\in\Csc$ be given by
$$\hat c_0(p)=\sup_{\varphi\in B_0(\Sigma)}\left(\hat I_0(\varphi)-\int\varphi\,dp\right)\quad\mbox{ for all }p\in\Delta.$$
Then $\hat c_0\apru c_0$. By the definition of $B^*$ and \eqref{eq:fen:dual}, we have
\begin{align*}
b\in B^*\Leftrightarrow &\max_{p\in\Delta}\left(\int\varphi\,dp-b(p)\right)\ge I_0(\varphi)\mbox{ for all }\varphi\in B_0(\Sigma,u(X))\\
\Rightarrow
 &\max_{p\in\Delta}\left(\int\varphi\,dp-b(p)\right)\ge \hat I_0(\varphi)\mbox{ for all }\varphi\in B_0(\Sigma)\\
\Rightarrow &\max_{p\in\Delta}\left(\int\varphi\,dp-b(p)\right)\ge \min_{p\in\Delta}\left(\int\varphi\,dp+\hat c_0(p)\right)\mbox{ for all }\varphi\in B_0(\Sigma)\\
\Rightarrow &\inf_{\varphi\in B_0(\Sigma)
}\left\{\max_{p\in\Delta}\left(\int\varphi\,dp-b(p)\right)- \min_{p\in\Delta}\left(\int\varphi\,dp+\hat c_0(p)\right)\right\}\ge0\\
\Rightarrow &\min_{p\in\Delta}[(b(p)+\hat c_0(p)]\le0,
\end{align*}
which implies the ``$\subseteq$" part of \eqref{eq:B*:vpa}. 
Conversely, the ``$\supseteq$" part of \eqref{eq:B*:vpa} is obvious. Therefore, \eqref{eq:B*:vpa} is proved.\qed

\subsection{Proof of Proposition \ref{prop:averse:nive}}
\begin{description}
\item[(i)$\Rightarrow$(ii)] Assume $\succsim$ is more ambiguity averse that an SEU preference with prior $p_0\in\Delta$. Then by Propositions  \ref{prop:comp:nive} and \ref{prop:ibp:nive} and Example \ref{ex:meu:seu}, we have, for any $c\in C^*$,
$c\le\delta_P$ for some convex and compact subset $P\subseteq\Delta$ containing $p_0$, which implies $c(p_0)\le0$. This completes (i)$\Rightarrow$(ii).

\item[(i)$\Rightarrow$(iii)] Assume $\succsim$ is more ambiguity averse that an SEU preference with prior $p_0\in\Delta$. Then by Propositions  \ref{prop:comp:nive} and \ref{prop:ibp:nive} and \eqref{eq:seup0}, we have, for any convex and compact subset $P\subseteq\Delta$ containing $p_0$,
that $\delta_P\in B^*$. Particularly, $\delta_{\{p_0\}}\in B^*$.
\end{description}

The proof of (ii)/(iii)$\Rightarrow$(i) is easy.\qed

\newpage

\section*{References}{
\begin{description}
\itemsep=0pt
\parskip=0pt

\item Amarante, M. (2009): ``Foundations of Neo-Bayesian Statistics," \textit{Journal of Economic Theory} 144, 2146--2173.

\item Anscombe, F.J., and R.J. Aumann (1963): ``A Definition of Subject Probability," \textit{The Annals of
Mathematical Statistics} 34, 199--205.

\item Arrow, K.J., and L. Hurwicz. (1972): ``An Optimality Criterion for Decision Making under Ignorance," In C.F. Carter and J.L. Ford, \textit{Uncertainty and Expectations in Economics}. Basil Blackwell: Oxford.

\item Cerreia-Vioglio, S., P. Ghirardato, F. Maccheroni, M. Marinacci, and M. Siniscalchi (2011): ``Rational Preferences under Ambiguity," \textit{Economy Theory} 48, 341--375.

\item Cerreia-Vioglio, S., F. Maccheroni, M. Marinacci, and L. Montrucchio (2011):
``Uncertainty Averse Preferences," \textit{Journal of Economic
Theory} 146, 1275--1330.

\item Cerreia-Vioglio, S., F. Maccheroni, M. Marinacci, and A. Rustichini (2014): ``Niveloids and Their Extensions: Risk Measures on Small Domains,"
\textit{Journal of Mathematical Analysis and Applications} 413, 343--360.

\item Dolecki, S., and G.H. Greco (1995):
``Niveloids," \textit{Topological Methods in Nonlinear Analysis} 5,
1--22.

\item Dunford, N., and J.T. Schwartz (1958):
\textit{Linear Operators Part I: General Theory}.
New York: Interscience Publishers.

\item Ellsberg, D. (1961):
``Risk, Ambiguity, and the Savage Axioms," \textit{Quarterly Journal
of Economics} 75, 643--669.

\item F\"ollmer, H. and A. Schied (2016): \textit{Stochastic
Finance: An Introduction in Discrete Time (4th Edition)}. 1st Edition: 2002. Berlin:
Walter de Gruyter.

\item Ghirardato, P., F. Maccheroni, and M. Marinacci (2004):
``Differentiating Ambiguity and Ambiguity Attitude," \textit{Journal
of Economic Theory} 118, 133--173.

\item Ghirardato, P., and M. Marinacci (2001):
``Risk, Ambiguity and the Separation of Utility and Beliefs,"
\textit{Mathematics of Operations Research} 26, 864--890.

\item Ghirardato, P., and M. Marinacci (2002):
``Ambiguity Made Precise: A Comparative Foundation," \textit{Journal
of Economic Theory} 102, 251--289.

\item Gilboa, I., and D. Schmeidler (1989): ``Maxmin Expected Utility with Non-Unique Prior," \textit{Journal of Mathematical Economics} 18, 141--153.

\item Giraud, R. (2005): ``Objective Imprecise Probabilistic Information, Second Order Beliefs and Ambiguity Aversion: an Axiomatization," 4th International Symposium on Imprecise Probabilities and Their Applications. Pittsburgh, Pennsylvania.

\item Heath, C. and A. Tversky (1991): ``Preference and belief: Ambiguity and competence in choice under uncertainty,"
\textit{J. Risk Uncertainty} \textbf{4}, 5--28.

\item Hansen, L., and T. Sargent (2000):
``Wanting Robustness in Macroeconomics,"
Mimeo, University of Chicago and Stanford University.

\item Hansen, L., and T. Sargent (2001):
``Robust Control and Model Uncertainty," \textit{American Economic
Review} 91, 60--66.

\item Hurwicz, L. (1951): ``Some Specification Problems and Application to Econometric Models," \textit{Econometrica} \textbf{19},  343--344.

\item Jia, G., J. Xia and R. Zhao (2020): ``Monetary Risk Measures," working paper.

\item Maccheroni, F., M. Marinacci, and A. Rustichini (2006):
``Ambiguity Aversion, Robustness, and the Variational Representation
of Preferences," \textit{Econometrica} 74, 1447--1498.

\item Mao, T., and R. Wang (2020): ``Risk aversion in regulatory capital principles," \textit{SIAM Journal on Financial Mathematics} 11, 169--200.

\item Mertens, J.-F., S. Sorin, and S. Zamir (2015): \textit{Repeated Games}. New York: Cambridge University Press.

\item Savage, L.J. (1954):
\textit{The Foundations of Statistics}.
New York: Wiley.
Second edition published by Dover in 1972.

\item Schmeidler, D. (1989):
``Subject Probability and Expected Utility without Additivity,"
\textit{Econometrica} 57, 571--587.

\item Strzalecki, T. (2011):
``Axiomatic Foundations of Multiplier Preferences,"
\textit{Econometrica} 79, 47--73.

\end{description}

\end{document}